**Title:** Description of a Differential Setup for Relaxation Microcalorimetry

**Authors**: J.V. Leitão[a]*, P. van Dommelen[a], F. Naastepad[a], E. Brück[a]

(a) Delft University of Technology, Section of Fundamental Aspects of Materials and Energy,

Mekelweg 15

Delft, 2619 JB, Netherlands

* Corresponding author: J.C.VieiraLeitao@tudelft.nl

Tel.: +31 (0)152789753, Fax: +31 (0)152788303

**ABSTRACT**

A specific heat measuring instrument, with the capacity for the application of magnetic fields up to 9 Tesla, resorting to microcalorimetry chips from the company Xensor Integration, has been successfully assembled and its functioning specifications are reported in the current paper.
With this instrument is it possible to perform specific heat measurements with applied magnetic fields up to 9 Tesla in milligram samples. This offers our group the possibility to calculate the actual adiabatic temperature change of a material, as well as providing reliable and precise information on any phase transition that may be influenced by the application of a magnetic field.

# 1 - INTRODUCTION

Since the beginning of magnetocaloric research, technically inaugurated by Brown in 1978 (Gschneidner and Pecharsky, 2008), the method most generally used for the evaluation of the magnetocaloric potential of a given material is the application of Maxwell's thermodynamic equation to a set of appropriately measured magnetic isotherms, so as to calculate the magnetic entropy change (Debye, 1926). Controversies regarding the validity of this equation aside, the usual alternative to this method is the calculation of the adiabatic temperature change, which translates the actual cooling and heating of a material.

The calculation of this temperature change relies on isofield specific heat measurements, as well as the same magnetization measurements as required by Maxwell's thermodynamic equation mentioned above (Tishin and Spichkin, 2003). Unfortunately this calculation is usually made difficult by the lack of a commercial measurement system that would allow for reliable specific heat measurements with an applied magnetic field.

Typically, to overcome this issue, many research groups world wide resort to assembling their own isofield specific heat measuring equipments. Among the most recent examples we may cite the setup described by Marcos (2003), consisting of an insert that can be fitted to any cryostat with the capacity to generate a magnetic field. This setup resorts to thermobatteries which give a voltage output in response to the heat exchange with the measured sample, which can then be read and interpreted as numerical value.

One other example may be observed in Korolev (2005), although this system has been specifically designed to measure magnetic colloids. Instead of using a permanent magnet it is designed as a microcalorimetry cell placed between the two poles of an electromagnet so as to generate a (low intensity) magnetic field.

The setup described by Kuepferling (2007) on the other hand resorts to commercial Peltier cells, a thermoelectric device made of a series of junctions of conductors with different thermoelectric power, acting as both sensors and actuators, being in this way able to achieve strict isothermal conditions. However this setup is highly dependent on accurate calibration of the Peltier cells.

Under this perspective the microcalorimetry chips from the company Xensor Integration have gained increasing relevance, due to their precision, practicality and relatively small price, as exemplified by Morrison (2008) or Minakov (2005a, 2005b, 2007), making them a very attractive and promising tool for the planning and assembly of such calorimeters.

We now report the design and construction of an experimental setup that would allow for specific heat measurements under high magnetic fields, resorting to such microcalorimetry chips. We have adopted a two chip setup in our equipment which enables us to easily bypass many bothersome calibration and equipment specific issues. This instrument's potential ranges well beyond the purely magnetocaloric oriented, as it can provide invaluable information regarding any phase transition where the application of a magnetic field may play a significant role.

**2 - EXPERIMENTAL SETUP**

**2.1 - Cryostat**

For the base of this setup we used a commercial cryostat from American Magnetics Inc. (AMI), equipped with a 9 Tesla 2 inch bore superconducting magnet (Solenoid), its own power supply and field programmer.

This cryostat has a 36 liter LHe reservoir, in direct contact with the superconducting magnet so as to keep it at a constant temperature of 4.2 K. A separate $LN_2$ reservoir, also with a capacity for 36 liters, is also present so as to reduce Helium evaporation.

The original Variable Temperature Insert (VTI), fitted for transport measurements, which originally came with this equipment was removed, so as another one could be fashioned with the capacity to perform specific heat measurements. As a consequence this made it impossible to use the built-in temperature control system of this cryostat.

A schematic diagram of the cryostat is shown in Figure 1.

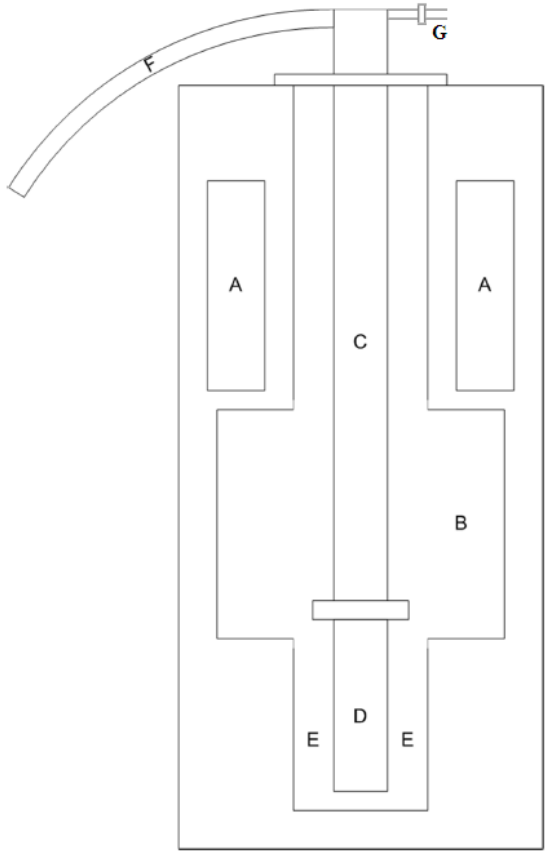

**Figure 1 - Schematic diagram of the AMI cryostat. Legend: A) Liquid Nitrogen reservoir; B) Liquid Helium reservoir; C) Insert; D) Magnetic field center; E) Superconducting Magnetic; F) Wiring conecting the insert to the rest of the measurement equipment; G) Vacuum pump;**

The evacuated sample space is in direct contact with the liquid Helium and consists of a small 25 mm wide cylinder protected by three cylindrical shields. The first of these shields is equipped with a heater. A scheme of the insert head in shown in Figure 2:

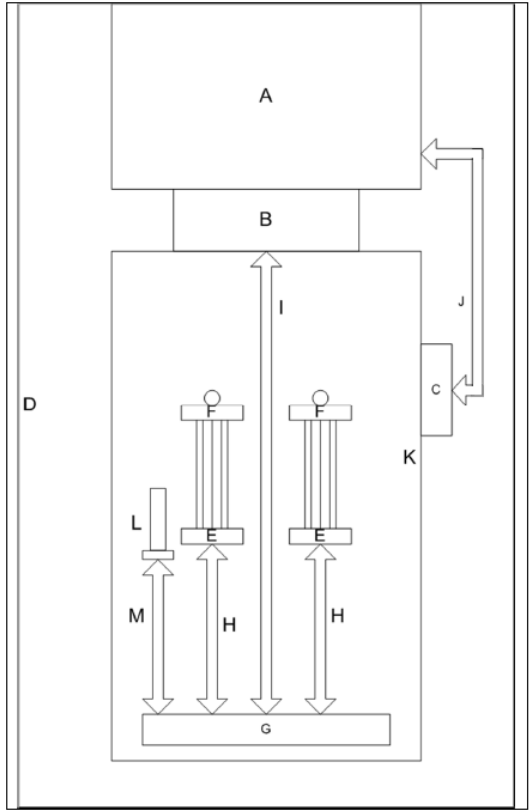

**Figure 2 - Insert head scheme. Legend: A) Body of the insert; B) Connector, enabling the head to be detached from the insert body and easily handeled; C) Wiring connection to the heating cylinder; D) Vacuum cylinder; E) Calorimetry chip socket; F) Xensor microcalorimetry chips (see below); G) Soldering platform functioning as a thermal anchor for all the wiring inside the insert head; H) Wiring between the chip sockets and the soldering platform; I) Wiring from the soldering platform to the body of the insert; J) Wiring from the heating cylinder to the body of the insert; K) Heating cylinder; L) Carbon glass temperature dependent resistor; M) Wiring from the resistor to the soldering platform;**

## 2.2 - Temperature Control

All aspects of temperature control in the current setup are done by a Lakeshore 331 Temperature Controller. This instrument measures the resistance of the carbon glass resistor, placed next to the calorimetry chips and in thermal contact with the heating shield, using a 4-point measurement method. This resistance value is then converted into a temperature through a calibration table previously measured. Furthermore, this temperature controller determines and supplies current to the heater by comparing the measured temperature with a set point temperature. As cooling is achieved passively by the simple contact of the insert head with the LHe, this heater is the only active element in temperature control.

## 2.3 - Microcalorimetry Chips

As the use of strong magnetic fields in metallic samples can give rise to high magnetization gradients that could possibly affect the measurement by moving or dislocating samples, we have decided to use small samples in the order of micrograms so as to minimize this effect. For this end we resort to two XEN-39328 microcalorimetry chips manufactured by the company Xensor Integration.

These chips consist of a thin 0.9x0.9 mm SiN membrane (Herwaarden, 2010) with a sensitive thermopile and a heater. Given their small size, these chips have a high thermopile and heater accuracy.

Chip specifications are displayed in Table 1.

**Table 1 - XEN-39328 chip manufacturer's specifications at 22 °C. These chips are largely similar to the older model XEN-39287.**

| | |
|---|---|
| Membrane dimensions | 0.9x0.9 mm |
| Approximate thermopile sensitivity | 2.0 mV K$^{-1}$ |
| Heating site dimensions (hotspot) | 92x92 µm |
| Pins | TO-5 |
| Heater resistance | 1.2 kΩ |
| Heater resistance temperature coefficient | 0.1 % K$^{-1}$ |
| Effective heat capacity (in air) | 100 nJ K$^{-1}$ |
| Maximum heating voltage (in vacuum) | 2.7 V |
| Membrane thermal resistance | 50 – 100 kK W$^{-1}$ |
| Membrane thermal resistance temperature | 0 % K$^{-1}$ |

coefficient

Thermopile resistance                                30 kΩ

As is shown in Figure 2, in the current setup we use two of these chips, one loaded with a Cu reference and the other with our sample.

This approach, contrasting with the single chip setups used by both Minakov (2005b) and Morrison (2008), was chosen so as to minimize any chip dependent issues that might interfere with our measurements. Given that the two chips used are equal in all aspects, and by replicating the same conditions in both of them, we can thus use the heat capacity ration between them and disregard most of the chips' influence on the measured results, making our subsequent data analysis simpler. In particular, this approach eliminates the necessity of having to deal with the unknown Seebek coefficient of the chip thermopiles (Morrison, 2008), the non constant resistance of the chip heaters with temperature, and other bothersome calibrations required for temperature scanning experiments.

To insure good thermal contact between the heaters on the chips, their sensors and the sample, Apiezon N and H greases are used, for low and high temperature measurements respectively. This further prevents sample motion during chip handling.

Apiezon N grease's specific heat has been extensively studied (Bunting *et al.*, 1969, Schnelle *et al.*, 1999) and as such can easily be taken into consideration during data analysis; this issue is further reduced due to the double chip setup, given that the amount of grease in both chips is comparable.

**2.4 - External Measurement and Power Supply Equipment**

The power supplied to the chips comes from a Keithley 2400 Source Meter. To measure and register the voltage output of the calorimetry chips' thermopiles a Keithley 2000 is used. This is equipped with a scanner card, enabling us to switch between different channels and measure different signals, thus simultaneously measuring both the reference and the sample chip.

The information supplied by the previous two modules and the Lakeshore temperature controller is then fed into a PC, via a GPIB bus, where the data is processed and analyzed via a battery of LabVIEW and MATLAB programs and routines.

## 3 - MEASUREMENT CYCLE

The specific heat can be determined by the so-called relaxation method. In this method, initially the sample and the chip are at the same temperature, $T_1$. A current is then applied to the chip's heater, making the sample temperature rise to a new constant, $T_2$. Next, the current is cut and the sample relaxes back to its original temperature, in a relaxation time of $\tau$ (Kim-Ngan, 1993). This cycle is achieved in this setup by supplying the chip heaters with a square wave from the Keithley 2400, and then, via the chips' inbuilt thermopiles, measuring the temperature of the sample and reference alternatively with the Keithley 2000, as to calculate one data point we need to measure both chips. Figure 3 shows the typical response to one relaxation period.

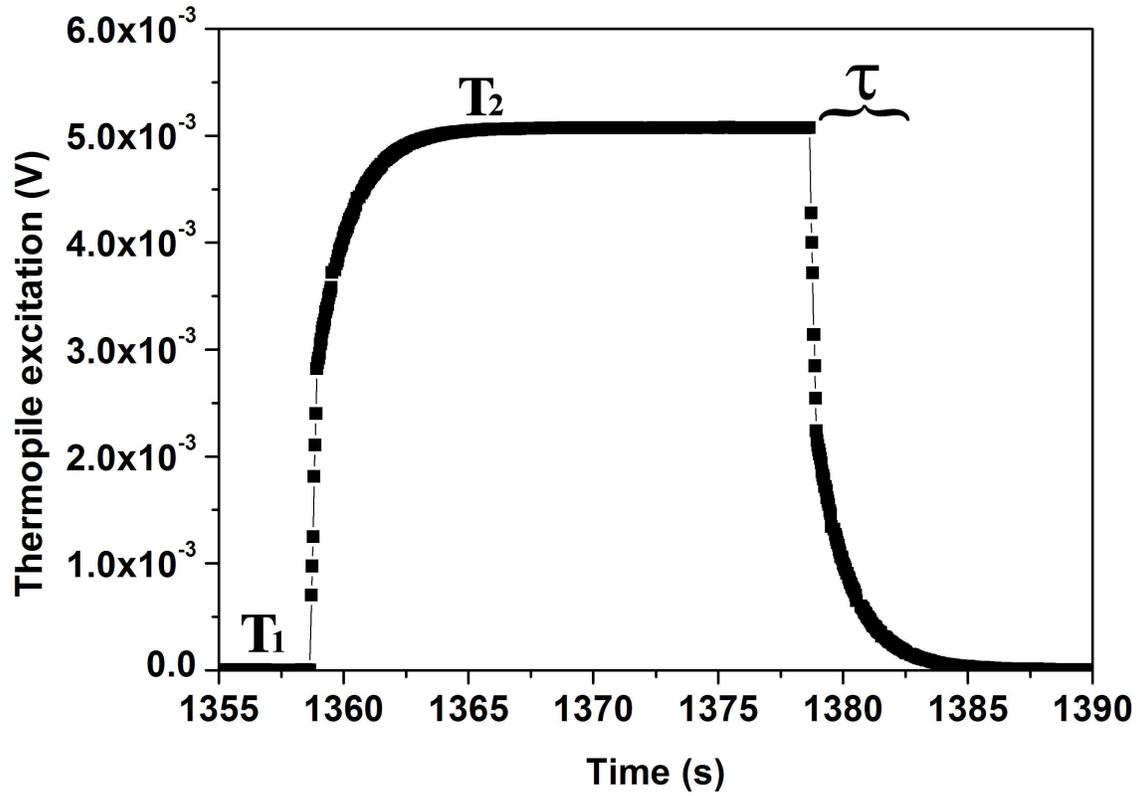

**Figure 3 - Thermopile response to the square wave heater input used to perform a relaxation measurement;**

The downward slope of one such measurement has the form given in eq. 1.

$$V(t) = A \cdot e^{-\frac{t}{\tau}} \qquad (1)$$

The mentioned relaxation time is then calculated by taking the logarithm of the signal and performing a least-square fit, as the slope of this fit is equal to the inverse of the relaxation time $\tau$. We can then calculate the heat capacity by the simple expression:

$$C = \kappa \tau \qquad (2)$$

Where $\kappa$, the thermal conductivity, is

$$\kappa = \frac{P}{T_2 - T_1} \qquad (3)$$

and $P$ is the power supplied to the chip.

## 4 - DATA MANAGEMENT

A complete measurement taken on this set up is depicted in Figure 4.

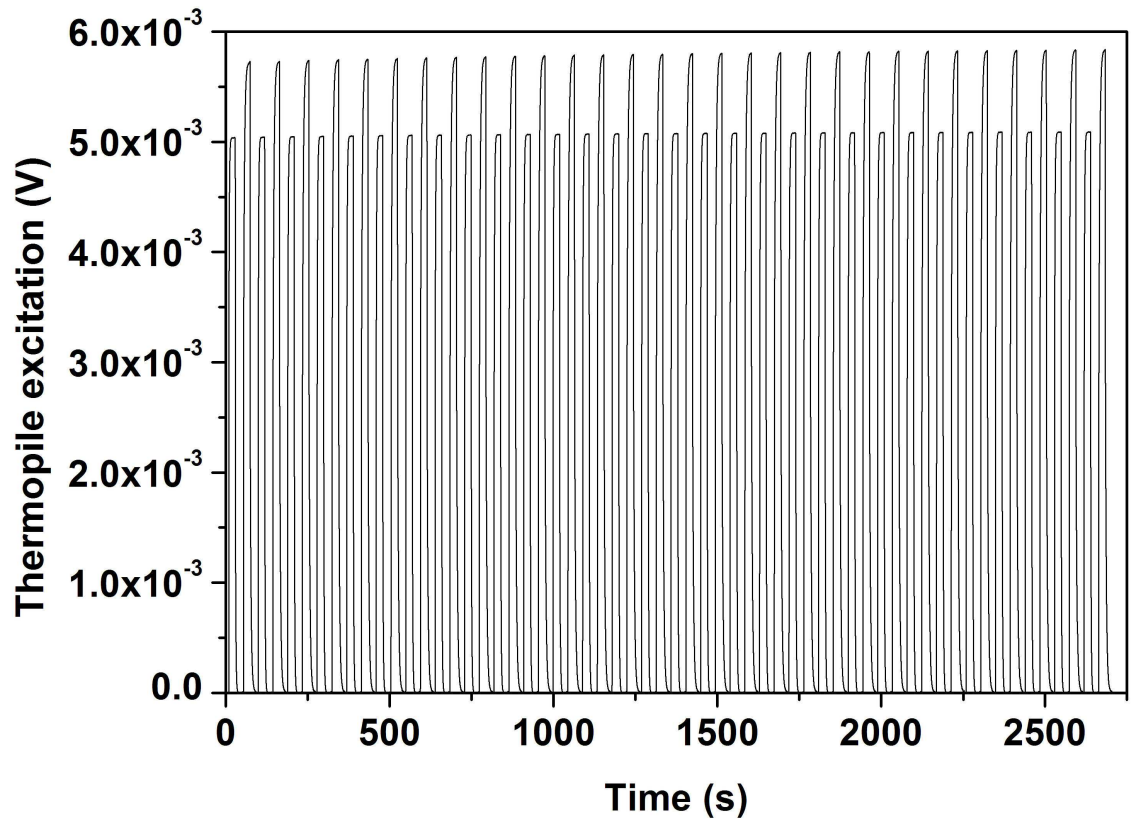

**Figure 4 - Specific heat data as aquired by the Keithley 2000. The alternative hight of the relaxation cycles is due to the equipment alternating between the sample and the reference chip.**

The calculated relaxation slope is shown in Figure 5. This calculation is done by trial and error, as certain particularities of the measurement system need to be taken into consideration, such as a systematic instantaneous temperature jump in the sensor temperature of the microcalorimetry chips. To correct for this problem in particular we ignore the first points of the fit.

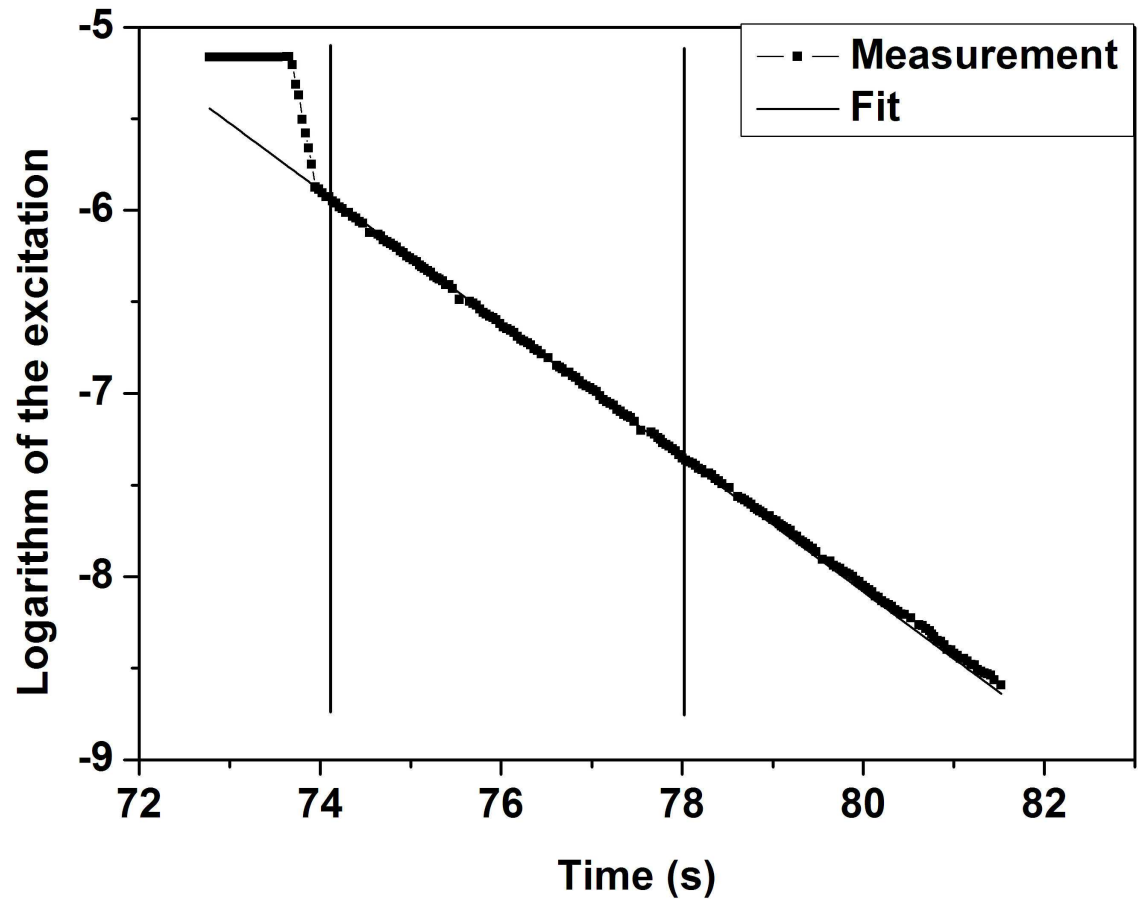

**Figure 5 - Linear fit of the logarithm of the thermopile output while relaxing.**

From the calculated relaxation time and the initial voltage supplied by the Keithley 2400 the ratio of the total heat capacity between the two chips can then be calculated, as displayed in Figure 6.

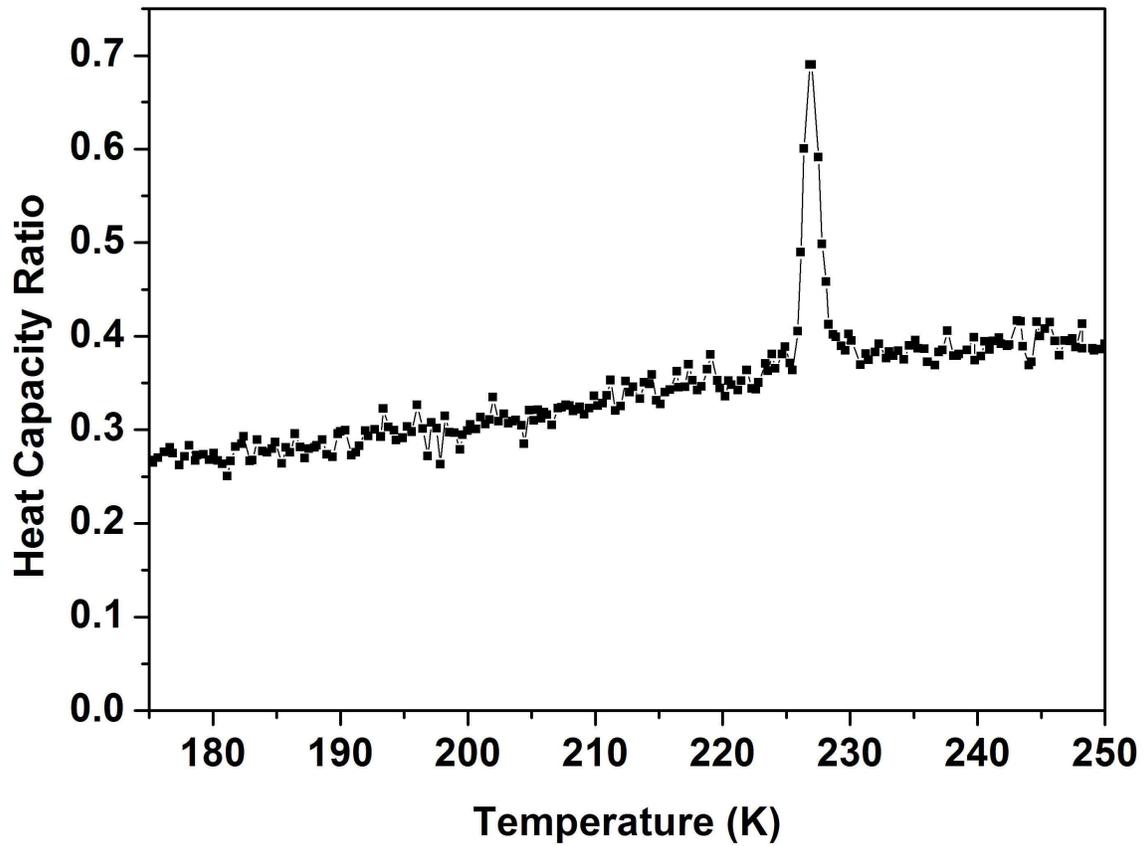

**Figure 6** - Ratio of sample and reference heat capacity over time. In the current case a Cu and a polycrystalline $Fe_2P$ sample were used as reference and sample respectively.

For this early measurement the reference used was a Cu piece and the sample measured a polycrystalline $Fe_2P$ piece. Their details can be consulted in Table 2:

**Table 2** – Details regarding the samples in the two chips used in the measurement displayed in Figure 6.

|  | Sample Mass (mg) | Mass error (mg) | Grease mass (mg) | Mass error (mg) |
| --- | --- | --- | --- | --- |
| Sample | 0,22 | 0,02 | 0,00 | 0,02 |
| Reference | 0,35 | 0,02 | 0,00 | 0,02 |

## 5 - Conclusion

A method for measuring the heat capacity of microgram samples has been successfully assembled using microcalorimetry chips from the company Xensor Integration. Even though this is not demonstrated in the current paper, with this calorimeter it is possible to measure the specific heat capacities of samples in a magnetic field and at various temperatures as well as the entropy change during magnetic field sweeps.

For this end we use a relaxation method measurement where the heat capacity is calculated from the sample relaxation time after the submission of the chip's heater to a square wave.

This setup uses two microcalorimetry chips simultaneously so as to eliminate the need for thermopile calibration and concerns regarding the Seebeck coefficient in magnetic fields and other equipment particularities.

**Acknowledgments**

The authors wish to acknowledge the funding from BASF Future Business and FOM (Stichting voor Fundamenteel Onderzoek der Materie), under the Industrial Partnership Programme IPP I18 of the 'Stichting voor Fundamenteel Onderzoek der Materie (FOM)' which is financially supported by the 'Nederlandse Organisatie voor Wetenschappelijk Onderzoek (NWO)', to the accomplishment of the research presented in this paper.